\documentclass[12pt]{article}
\usepackage{psfig}
\newcommand{\beq}{\begin{equation}}
\newcommand{\eeq}{\end{equation}}
\newcommand{\bea}{\begin{eqnarray}}
\newcommand{\eea}{\end{eqnarray}}
\newcommand{\tev}{\rm TeV}
\newcommand{\drawsquare}[2]{\hbox{%
\rule{#2pt}{#1pt}\hskip-#2pt
\rule{#1pt}{#2pt}\hskip-#1pt
\rule[#1pt]{#1pt}{#2pt}}\rule[#1pt]{#2pt}{#2pt}\hskip-#2pt
\rule{#2pt}{#1pt}}

\newcommand{\Yfund}{\raisebox{-.5pt}{\drawsquare{6.5}{0.4}}}

\begin{document}

\thispagestyle{empty}
\begin{flushright}
UW/PT-02-17\\ HUTP-02/A039\\
hep-th/0208120\\ \today
\end{flushright}
\vspace{1mm}
\begin{center}
\vspace*{1cm}
{\Large \bf Deconstruction and Gauge Theories in AdS$_5$}\\
\vspace*{0.8cm}
{\large Lisa Randall$^a$, 
Yael Shadmi$^b$\footnote{Incumbent of a Technion Management Career
Development Chair}
and 
Neal Weiner$^c$}\\
\vspace{1.5mm}
\vspace{0.8cm}
$^a$ {\em
Jefferson Physical Laboratory, Harvard University, Cambridge, MA 02138}\\
$^b$ {\em Physics Department, Technion, Haifa 32000, Israel}\\
$^c$ {\em Department of Physics, University of Washington, Seattle, WA, 
98195}
\vskip 0.2in
\vspace{0.7cm}
ABSTRACT
\end{center}
On a slice of AdS$_5$, despite having a dimensionful coupling, gauge theories can exhibit logarithmic dependence on scale.
In this paper, we utilize 
deconstruction to analyze the scaling behavior of the theory, both above and below the AdS curvature scale, and shed 
light on position-dependent regularizations of the theory. We comment on 
applications to geometries other than AdS.

\setcounter{page}{0} \setcounter{footnote}{1}
\newpage

\section{Introduction}
In recent years, it has become clear that extra dimensions can play an 
important role in our understanding of the hierarchy problem 
\cite{Arkani-Hamed:1998rs,Randall:1999ee}. 
In \cite{Randall:1999ee}, it 
was noted that the cutoff scale of the standard model could be 
$O(\tev)$, and yet have weakly coupled gravity if the standard model 
fields were localized on a brane on the boundary of a slice of AdS$_5$.

One seeming consequence of having a cutoff at a TeV is that it is 
impossible to discuss issues of the UV, such as high scale unification. However, 
this apparent obstacle no longer applies when the gauge sector of the standard model is 
placed in the bulk of AdS. In particular,  in \cite{Pomarol:2000hp} the couplings of fields on the UV brane were studied, and it was demonstrated explicitly that
despite the dimensionful bulk gauge coupling, the 
effective coupling of the zero mode runs logarithmically below the AdS curvature scale. In \cite{Randall:2001gc,
Randall:2001gb}, a position-dependent regulator motivated by holography was used to show that the predictions of perturbative high-scale unification were possible, even in theories where standard model fields experience a strong-coupling cutoff at the TeV scale. More recently, 
\cite{Goldberger:2002cz,Agashe:2002bx,Choi:2002zi,Contino:2002kc,Goldberger:2002hb,Choi:2002ps} have studied the behavior of gauge fields in AdS. 
All of these works agree that, at some level, gauge running in AdS is 
logarithmic, rather than power law, below the compactification scale. 
However, there nonetheless remains some confusion as to what constitutes a 
sensible regulator, and in what sense one can talk about a zero mode 
definitively above the IR brane scale.

Over the past year, 
``deconstruction''~\cite{Arkani-Hamed:2001ca,Hill:2000mu} 
has a appeared as a new tool for model 
building and understanding the properties of higher-dimensional 
gauge theories. By viewing the higher-dimensional theory as a product of 
four-dimensional theories, one can easily study the properties of the 
theory at intermediate energies. In this paper, we will employ 
deconstruction to give us an understanding of the properties of AdS 
gauge theories at energies above the scale of the IR 
brane.\footnote{Deconstructing gauge theories in RS was also
discussed in~\cite{Cheng:2001nh,Abe:2002rj}.}
We, too, 
will recover log running above a TeV, but we will also see how position 
dependent regulators naturally arise in the deconstructed framework. We also will recover cutoff sensitive ``power-law running'' effects above the curvature scale, which will appear as threshold effects associated with the cutoff of the theory. 
We will also see in what sense we can discuss a renormalized zero mode 
above the IR brane scale. Lastly, the results we achieve can be straightforwardly applied to different geometries.

\section{Gauge Theories in AdS}
Five dimensional gauge field theories are defined as cutoff theories. Because 
the gauge coupling is dimensionful, there is an associated ultra-violet 
strong coupling scale. However, AdS gauge theories present a number of 
subtleties not found in flat extra dimensions.

In flat extra dimensions, we see the first Kaluza-Klein mode at a scale 
$1/R$. At a scattering energy $E \simeq n/R$, one exchanges $n$ KK 
modes, giving a strong coupling scale $\Lambda  \sim 4 \pi/ R g_4^2 \sim 
4 \pi^2/g_5^2$. Note that we could live on a brane at any point in the 
space and this formula would remain essentially unchanged.

In AdS, the situation is quite different. If we consider the RS1 
space with UV and IR cutoff branes, the mass of the first KK mode is set 
by the IR scale as is the spacing between KK modes. Thus, a scattering 
at a high scale $\Lambda \gg M_{IR}$ would naively be strongly coupled. 
The essential difference in AdS is that this is a position dependent 
question. In the language of KK modes, this is because the light KK 
modes are strongly coupled to the IR brane, but very weakly coupled to 
the UV brane.

This result is not surprising because it arises from the basic 
symmetries of AdS. Let us consider a $U(1)$ gauge theory in AdS. We can 
write the metric as
\beq\label{metric}
ds^2= e^{-2\sigma(y)} dx^4 -dy^2\ .
\eeq
For now and unless otherwise noted, we take $\sigma(y)=k y$.
With this metric, the gauge boson action is
\begin{equation}\label{fived}
S= - {1\over 4}\,\int d^4x dy
{1\over g_5^2}\, \left[G_{\mu\nu}^2 -2e^{-2\sigma}
  G_{\mu 5}^2\right]\ ,
\end{equation}
where all 4d contractions are with $\eta_{\mu\nu}$. Note that this 
action is invariant under the transformation
\bea \label{transform}
y\rightarrow y+ \delta, \\
x\rightarrow e^{k \delta} x, \\
\partial_\mu \rightarrow e^{-k \delta} \partial_\mu, \\
A_\mu \rightarrow e^{-k \delta} A_\mu.
\eea
The transformation of the gauge field is necessary to insure  that 
gauge covariant derivatives of charged fields are also covariant under 
the rescaling.

This symmetry tells us that a process with four-momentum $p_\mu$ at a 
point $y=y_0$ relates to an equivalent process with four-momentum $e^
{-k \delta} p_\mu$ at a point $y'=y_0+\delta$. Or (not surprisingly) 
the local (in $y$) strong coupling scale scales with the warp factor.

In going to the quantum theory, we have to be careful in how we define 
our couplings. In four-dimensional QED, for instance, we define the 
coupling $e$ at a scale $\mu_0$, with the same $e$ and the same $\mu_0$ 
everywhere. This is a requirement for preserving the Poincar\'e invariance 
of the theory. In AdS we do not have five-dimensional Poincar\'e 
invariance so we must be careful to define our theory to preserve the 
symmetries.

One way we {\em cannot} define the theory is to specify all couplings at the 
scale of the UV brane. This is because over most of the space the gauge 
theory becomes strongly coupled below this scale, and hence this is not even a region in which the perturbative field theory is defined. 

Likewise, we could specify the couplings
at the scale of the IR brane, 
but this is troubling as we take the IR brane to infinity, and, 
moreover, it would spoil the symmetries of the theory. Hence, the most 
natural definition, the one that maintains the symmetry of the space, is 
to assume that the couplings are specified at a scale which scales with the 
warp factor in $y$. This will be essential in writing down the 
deconstructed quantum theory.

\section{The classical deconstructed theory}
Let us begin by reviewing the standard process of deconstruction. Our goal will be to write a five-dimensional  $SU(n)$ as a four-dimensional, $SU(n)^N$ gauge 
theory with
bifundamental fields, $Q_i$ (see table~1) with\footnote{We will comment on the supersymmetric version of the theory at
end of this section.}
\beq
\langle Q_i\rangle=v_i\, {\rm diag} (1,1,\cdots,1).
\eeq 
The lattice will more accurately reproduce intermediate scale physics with increasing $N$, the number of lattice sites. Most importantly, the theory is cut off at a local scale $\Lambda_j \sim v_j$, so that we can regulate our theory in a gauge-invariant fashion.

From the outset, deconstruction of an AdS gauge theory is quite 
different from the deconstruction of one in flat space. In flat space, 
if a fermion is localized at any point in the additional dimension(s), 
it will couple with comparable strength to each of the KK modes. 
Consequently, accurately reproducing the intermediate energy features in 
the deconstructed theory requires matching states and energies.

The AdS case is quite different. Above the local AdS curvature scale, the 
gauge theory quickly becomes strongly coupled. 
However, those 
KK modes are very weakly coupled to the UV brane, so in a deconstructed 
theory, these modes will be automatically grouped, depending on the 
coarseness of the latticization, into single modes whose effects will 
replace multiple modes in the extra dimensional theory. While this seems 
peculiar, it is precisely because those modes are so weakly coupled that 
this remains a true description of the features of the higher 
dimensional theory.

We begin by discretizing the 5th dimension in~(\ref{fived}). We have
\begin{equation}\label{disfived}
S= -{1\over4}\,\int d^4x \, \left[
\sum_j {a_j\over g_{5 j}^2} {G_{\mu\nu}^{(j)}}^2 -2
\sum_j { a_j e^{-2\sigma_j}\over g_{5 j}^2}\,
{G_{\mu 5}^{(j)}}^2\right]\ ,
\end{equation}
where $j$ denotes the lattice site,
$a_j$ is the lattice spacing near site $j$,
$a_j=y_{j+1}-y_j$ and $j=1\ldots N$, and
\beq\label{disgmu5}
G_{\mu5}^{(j)}= \partial_\mu A_5^{(j)} - {1\over a_j}\,
(A_\mu^{(j+1)}-A_\mu^{(j)})\ .
\eeq
To write down a lattice action we can replace $A_5$
by the link fields
\beq\label{link}
U= e^{i \int_{y_j}^{y_{j+1}} dy A_5(x,y)}\ ,
\eeq
so that for small lattice spacing,
\beq
D_\mu U\sim i ( a_j\partial_\mu A_5^j - A_\mu^j+ A_\mu^{j+1})\ .
\eeq

\begin{table}
$$
\begin{array}{c|ccccc}
   &SU(n)_1&SU(n)_2&SU(n)_3&\cdots&SU(n)_N \\ \hline
Q_1&\Yfund& \overline{\Yfund}&1&\cdots&1 \\
Q_2&1&\Yfund& \overline{\Yfund}&\cdots&1 \\
    &\vdots&\vdots&\vdots&\ddots&\vdots   \\
Q_{N-1}&1&1&1&\cdots&\overline{\Yfund}\\
\end{array}
$$
\label{tb:decontent}
\caption{Field content for a deconstructed $SU(n)$ theory.}
\end{table}

Let us now consider the action of the deconstructed theory, 
\begin{equation}\label{fourd}
S= \int d^4x \left[-{1\over4} \sum_j {1\over g_j^2} {F_{\mu\nu}^{(j)} }^2
+\sum_j {1\over g_j^2}\, D_\mu Q_j^\dagger D^\mu Q_j\right]\ .
\eeq
The peculiar normalization for the matter fields is convenient
since they will ultimately supply the 5th component of the gauge
fields. In the supersymmetric version of the theory, which we
will later consider,
this normalization allows for a holomorphic definition
of the length of the 5th dimension, since the gauge coupling
does not appear in the relation between this length and the VEVs.

Assuming all $Q$'s get VEVs we have $N-1$ $\sigma$-models,
coupled by the gauge interaction. 
Writing $Q_j= v_j {\tilde U}_j$, we choose the parameters of the deconstructed theory, $v_j$ ang $g_j$, so  that the action~(\ref{fourd}) agrees with the 5d lattice
action~(\ref{disfived}):
\begin{eqnarray}\label{map}
{1\over g_j^2}&=& {a_j\over g^2_{5 j}},\\\nonumber
{v_j^2\over g_j^2}&=& { e^{-2\sigma_j} \over
 a_j g^2_{5 j}}\ ,
\end{eqnarray}
so that,
\beq\label{eq:va}
v_j^2= { e^{-2\sigma_j} \over
 a_j^2 }\ .
\eeq
Here we see the effect of the AdS geometry. 
While a theory with five-dimensional Lorentz 
invariance naturally has  a deconstructed description with all VEVs, $v_j$, 
equal, here the $v_j$'s scale with the warp factor.
We also note that, with conventional normalization for the bifundamentals,
the gauge coupling would appear in~(\ref{eq:va}).

We thus have a ``dictionary'' between the 5d parameters ---
the gauge coupling, $g_{5 j}$,
and the lattice spacing $a_j$ --- and the parameters of the
deconstructed theory, $v_j$ and $g_j$.

One can consider different variations of this basic theory.
For example, in order to study bulk matter, we can add
matter fields for every $SU(n)$ factor.
In the supersymmetric version of the theory (with ${\cal N}=1$ 
supersymmetry),
one needs to add matter in order to cancel anomalies
on the two ends of the chain. A simple possibility is to add
fundamentals and antifundamentals as shown in table \ref{tb:susydecon}.
\begin{table}
$$
\begin{array}{c|ccccc}
   &SU(n)_1&SU(n)_2&SU(n)_3&\cdots&SU(n)_N \\ \hline
{\bar q}_{i=1\ldots n}&\overline{\Yfund}& 1&1&\cdots&1 \\
q_{a=1\ldots n}&1& 1&1&\cdots&\Yfund \\
\end{array}
$$
\label{tb:susydecon}
\caption{Adding vectorlike fields at the end sites cancels anomalies in the supersymmetric theory.}
\end{table}
These fields can get mass through non-renormalizable operators 
\cite{Csaki:2001em}, or through extra vectorlike matter \cite{Skiba:2002nx}.
However, they do not affect our
basic analysis and results.

\subsection{Couplings at intermediate energies in the deconstructed 
theory}
To get a sense of the behavior of the deconstructed theory, let us begin 
by considering the exchange of a bulk gauge field coupling to a field on 
the UV brane at energies $R^{-1} \gg q \gg \tev$. Being confined to the 
UV brane is equivalent to being charged under the gauge group $G_1$ in 
the deconstructed theory.

At very low energies, the entire theory is Higgsed to the diagonal. Then 
the two-point function in the Feynman-t'Hooft gauge for exchange of a 
bulk gauge boson between two fields on the UV brane is given simply 
by\footnote{We use normalization such that the gauge coupling is 
included in the propagator for clarity.}
\beq
G_q(0)\eta^{\mu \nu} \simeq-i \frac{g^2}{N}\frac{\eta^{\mu \nu}}{q^2}.
\eeq
At a higher energy  $q \gg \tev$, we have resolved the independent gauge 
groups Higgsed down below a scale q, and instead, the dominant 
contribution comes from the exchange of the diagonal gauge field of the 
groups which have been Higgsed down at energies above $q$. This behavior is shown in figure \ref{fig:posreg}.
\begin{figure}
\centerline{
\psfig{file=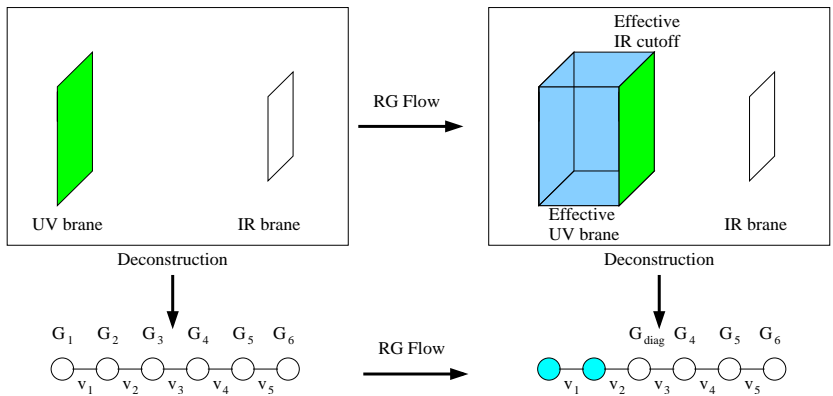,width=0.8\textwidth}
}
\caption{Connection between RG flow in the continuum and deconstructed 
theory. A high energy theory has the massive modes (localized near the 
UV brane) integrated out to leave a different theory with a different UV 
brane with large kinetic terms for the bulk gauge field. For momentum 
$q$ processes occuring on the UV brane, the dominant contribution to 
exchange comes from the diagonal mode of the shaded region, yielding an 
effective IR cutoff in the continuum theory.}
\label{fig:posreg}
\end{figure}
The effective number of 
groups contained in the diagonal at a scale $q$ can be written 
approximately as
\beq
a N_{eff} \simeq R \log (q/\Lambda).
\eeq
Now we can approximate the exchange of the gauge field by the diagonal 
mode, yielding
\beq
G_q(0) \eta^{\mu \nu} \approx -i \frac{g^2}{N_{eff}}\frac{\eta^{\mu 
\nu}}{q^2} \approx -i\frac{g_5^2 k}{\log (q/\Lambda) q^2}\,
\eta^{\mu \nu},
\eeq
where we have used the relation $g_5^2 = g^2 a$.

It is notable that the dominant classical logarithmic scaling for the AdS gauge field 
two-point function falls out automatically from the deconstructed theory, without invoking any further assumptions about the holographic nature of the theory. 
This is an excellent indication that the 
deconstructed theory is effectively reproducing the physics of the 
five-dimensional theory.

\subsection{Couplings at high energies in the deconstructed theory}
At high energies, our coarse latticization is insufficient. If we want 
to understand physics near the strong coupling scale of the theory, we 
should latticize such that each $\alpha_j = g_j^2/4 \pi \approx 4 \pi$. 
However in this case the separation of VEVs is small, that is, the $\sigma_j$'s are close together, and one would need 
to diagonalize the gauge boson mass matrix, and determine couplings to 
each mode. However, since the VEVs are close together, the result is 
obvious: groups Higgsed down significantly below the UV cutoff are 
irrelevant (classically), and groups Higgsed at a scale roughly an 
factor of a few down from the cutoff act nearly as a flat extra 
dimension with radius $R_{eff}\sim 1/k$, as expected.

\section{Gauge coupling running in deconstructed RS}
Before we can talk about running of the gauge coupling, we must define 
what we mean, precisely. We cannot speak perturbatively about fields on 
the TeV brane scattering at energies well above a TeV. In what sense can 
we discuss running?

The answer is actually quite straightforward. At low energies, 
fields on the TeV brane scatter through the zero mode of the 
five-dimensional theory. Likewise, fields on the UV brane scatter at low energies with 
equal couplings (as required by the remaining four dimensional gauge 
invariance). Thus, we can relate measured gauge couplings on the TeV 
brane to those that {\em would} exist for fields on the UV brane. We can 
then ask questions about two-point functions at higher energies on the 
UV brane.

The deconstructed theory is characterized
by two sets of parameters: the dimensionless couplings
$g_j$, and the dimensionful VEVs $v_j$.
As discussed above, in order to mimic
the continuum theory, we choose
\beq\label{defvk}
v_j=v e^{-j w} ,
\eeq
with $w= k a$.
In addition, the individual couplings are defined at the scale of
the corresponding VEVs, that is,
\beq\label{defg}
g_j=g_j(v_j)=g\ ,
\eeq
or, in terms of the strong coupling scales of the individual groups,
\beq\label{lambdas}
\Lambda_j^b = v_j^b \, e^{-{8\pi^2\over g^2}},
\eeq
where $b$ is the one-loop beta function coefficient,
which is the same for all groups.
Examining the relations~(\ref{map}), we see that this indeed corresponds
to taking
$g_{5 j}= const$ and $a_j=a$.\footnote{Note that the
same holds with conventional normalization
for the matter fields: Taking the VEVs to warp and defining the 
individual
couplings to equal a common value at {\it exponentially varying}
scales corresponds
to a 5d coupling and lattice spacing that have uniform values at {\it 
exponentially varying}
scales.}

It is instructive to start by considering a coarse lattice,
with the lattice spacing larger than the
AdS curvature: $a\gg k^{-1}$.
Then we have
\beq
v_1 \gg v_2\gg\cdots\gg v_{N-1}\ .
\eeq
This simplifies the analysis considerably:
we can study the running by turning on the VEVs one at a time.
It is amusing that the RS case turns out to be simpler
to analyze than the flat case, where all VEVs are equal
and therefore have to be considered together.

Let us now derive the coupling of the low energy theory.
At the scale $v_1$, $SU(n)_1\times SU(n)_2$ is broken to the diagonal
subgroup, which we label $SU(n)_{2 D}$ (with the diagonal group of $j$ groups labeled $SU(n)_{j D}$), with gauge coupling,
\beq
{1\over g_{2 D}^2(v_1)} = {1\over g_1^2(v_1)}+{1\over g_2^2(v_1)}\ .
\eeq
We can now run down to the scale $v_2$, where,
\beq
{1\over g_{2 D}^2(v_2)} = {1\over g_1^2(v_1)}+{1\over g_2^2(v_1)} +
{b\over 8\pi^2}\, \ln{v_2\over v_1}\ ,
\eeq
where we used the fact that the diagonal subgroups have the same beta function 
coefficients as the original group factors.

Repeating this $j$ times we find,
\beq\label{gl}
{1\over g_{j D}^2(v_j)} = {1\over g_1^2(v_1)}+\cdots +{1\over 
g_{j}^2(v_j)} +
{b\over 8\pi^2}\, \ln{v_j\over v_1}\ .
\eeq

Using the fact that $k=-w^{-1} \ln{v_j/v}$, and $g_5^2=a g^2$, this can 
be written as
\beq\label{gl1}
{1\over g_{D}^2(v_j)} = 
 - {1\over k g_5^2}\, \ln{v_j \over v} +
{b\over 8\pi^2} \ln{\mu \over v} - {b\over 8\pi^2} k a 
\eeq
Indeed, the low energy coupling depends logarithmically on the scale.
In fact, we see that the low energy coupling contains
two pieces with logarithmic energy dependence.
The first term on the RHS of eqn.~(\ref{gl1}) is a classical
piece, which reproduces the fact that the 4d and 5d couplings
are related by the ``radius''.
In the deconstructed theory, it is obtained by summing
over KK modes at tree level.
The second piece is a loop effect.
It is obtained from running the coupling between adjacent KK modes.

Let us now consider the supersymmetric theory. In the presence of ${\cal N}=1$
supersymmetry, holomorphy and symmetries dictate the dependence
of the diagonal $SU(n)$ strong coupling scale, $\Lambda_D$, on 
$\Lambda_j$, $v_j$, at any stage of the Higgsing.
In particular, at low energies,
\beq\label{lammatch}
\Lambda_D^b \sim {\Lambda_1^b\ldots \Lambda_N^b\over N^b  
v_1^b\ldots v_{N-1}^b} .
\label{eq:e26}
\eeq
The factor of $N^b$ is not dictated by symmetries, but can be found from explicit computation \cite{Csaki:2001zx}.\footnote{The factor $N^b$ can also be motivated by the following argument:
Consider the flat space theory. To properly reproduce the
continuum theory, the scale of the unbroken diagonal subgroup
should remain fixed as we take $N\rightarrow \infty$. Since
$v$ scales as $N$ (recall $v$ is the inverse lattice spacing),
and $g$ scales as $\sqrt{N}$ (so that the classical diagonal coupling
is fixed), the expression (\ref{eq:e26}) should contain the factor $N^b$.
}

The relation~(\ref{lammatch}) allows us to obtain the
low energy coupling for any profile of the VEVs.
In particular, for the RS geometry, with~(\ref{defvk}), (\ref{lambdas}),
we have at low energy
\beq
{1\over g_L^2(\mu)} = {N\over g^2} 
+ {b\over 8\pi^2}\ln{\mu\over v/N}.
\eeq
Running from a scale below $k$, deconstruction yields ordinary logarithmic running with a beta function set by the four-dimensional gauge theory. In terms of the parameters of the theory, this gives us
\beq
 \simeq
  - {1\over k g_5^2}\, \ln{v_{N-1} \over v} 
+{b\over 8\pi^2}\ln{\mu\over v/N} =
{R\over g_5^2} +  {b\over 8\pi^2}\ln{\mu\over v/N} .
\eeq
This indeed reproduces the expression we found for the
non-supersymmetric theory.
Again, we see the logarithmic dependence on the energy
scale in both the tree-level and one-loop pieces.

\subsection{Fine Latticizations and Power Law Running}
\label{sec:powerlaw}
So far we have seen that a coarse latticization of AdS can generate the log running behavior which is characteristic of these gauge theories. But what about energies above $k$? At these energies the space is approximately flat and there is a possible power law contribution between $k$ and $\Lambda$\footnote{We are indebted to Nima Arkani-Hamed for very useful discussions on these issues}.

This arises from a subtlety which we have to this point ignored. We have defined our couplings $g_j$ at the scale $v_j$. This was reasonable because the scales varied with energy, preserving the symmetries of the theory. Also, $v_j$ is approximately the local cutoff, so if we believe the unfication is only manifest in a microscopic theory (e.g. string theory), this is the approximately the cutoff scale (which is ambiguous).

These ambiguities make little difference when the coupling scales logarithmically with the cutoff, but is very important in the power law piece. To extract the behavior most straightforwardly, we will consider the supersymmetric theory, but the presence of the power law piece will not depend on the supersymmetric matter content. Consider again~(\ref{lammatch}). If we state that the couplings unify at a scale $M_j\sim v_j$ we have
\bea
\Lambda_D^b &\sim& {\Lambda_1^b\ldots \Lambda_N^b\over N^b  
v_1^b\ldots v_{N-1}^b},\\
\mu e^{8 \pi^2/g_D^2(\mu)} &\sim&  {\Lambda_1^b\ldots \Lambda_N^b\over N^b  
M_1^b\ldots M_{N-1}^b}\times \frac{M_1^b\ldots M_{N-1}^b}{v_1^b\ldots v_{N-1}^b},\\
&\sim& v^b e^{8 \pi^2 N/g^2} e^{\gamma N b},
\eea
where $\gamma = \log (M_j/v_j N^{1/N})$ is independent of $j$.

Restating this in terms of gauge couplings, we have
\beq
\frac{1}{g_D^2(\mu)} = \frac{N}{g^2} + \frac{b}{8 \pi^2} \log (v/\mu) + \frac{N b \gamma}{8 \pi^2}.
\label{eq:powerlaw1}
\eeq
Note that we have made no assumptions about the background geometry to derive this result - it can be applied to any space. In AdS we recognize the first term as the usual $R/g_5^2 = \log(v/v_{N-1})/k g_5^2$ piece, and the second we recognize as the log running piece. The final piece is the power law running piece. Using the relation
\beq
N=\frac{R}{a} = \frac{v \log(v/v_{N-1})}{k},
\eeq
we can rewrite~(\ref{eq:powerlaw1}) as
\beq
\frac{1}{g_D^2(\mu)} = \frac{N}{g^2} + \frac{b}{8 \pi^2} \log (v/\mu) + \frac{v b \gamma}{8 \pi^2 k} \log(v/v_{N-1}).
\label{eq:powerlaw2}
\eeq
Taking $v \sim \Lambda$, we see the linear cutoff dependence (``power law running'') explicitly. This term is worthy of two observations. The first is that it seems that merely by restating the each $g_j$ at $v_j$ we could remove the power dependence. It is true that a suitable redefinition of $g_5^2$ can absorb this piece entirely. However, since we are ultimately considering the possibility of unification, we cannot do this as the separate redefinitions of couplings for $3-2-1$ would not be $SU(5)$ symmetric. Hence, in considering {\em differences} of couplings, this is a real effect.

The second observation is that there is essentially a free parameter in the theory, $\gamma$. This is not surprising because the theory has a linear divergence, implying cutoff sensitivity. We have employed a particular cutoff which has an ambiguity: namely the unification scale relative to the cutoff scale. This uncertainty is reasonable from an effective field theory perspective as it merely parameterizes our ignorance of the physics which generates the low energy field theory. In terms of the continuum theory, the upshot is that while there is true power law dependence on the cutoff, the coefficient of this term is unknown. That being said, it should also be noted that the relative strengths of these cutoff terms are still proportional to the relevant $\beta$ functions.

Lastly, we note that here the power law piece can be subdominant to the log running piece.  We would expect additional threshold effects comparable to the power law piece in any complete model, which could easily spoil quantitative unification. In AdS, because the log piece can dominate for unification scales below the strong coupling cutoff of the theory, the GUT-violating threshold effects are under control and  one can still speak reliably about unification.

\section{Position Dependent Regularizations from Deconstruction}
In \cite{Randall:2001gb}, a position dependent cutoff was used to study 
the evolution of gauge couplings in a warped background. The 
regularization employed was an energy dependent IR brane, which 
was motivated by holography and the presence of strong 
coupling away from the UV brane at intermediate energy.
Here, this regularization arises naturally from the 
deconstructed theory, without  appealing to holography.

Let us begin by running the theory down from high energies. In the 
deconstructed theory, we gradually Higgs down more and more groups such 
that the highest energy group becomes more weakly coupled. At each level 
we integrate out the classical and quantum fluctuations of only the 
left-most groups in the quiver diagram. If we treat deconstruction as an 
invertible map, the resulting position space regularization is 
illustrated in figure \ref{fig:posreg}.

In the deconstructed picture the appropriate regularization is obvious. 
One merely integrates out the heavy modes, which exist near the UV 
brane, and replaces the theory with a different one, with a new UV brane 
positioned nearer to the IR, but now with large boundary kinetic terms 
corresponding to the modes integrated out.

Of course, this relates the fundamental Lagrangian with a lower energy 
effective theory. We also want to ask questions about the relationship 
between low-energy and high-energy scatterings on the UV brane. Here, 
our interpretation of the above diagram is somewhat different. At any 
given momentum $q$, we are cut off from the sites Higgsed at scales less 
than $q$. As a consequence, we should calculate loops only with the 
modes from the sites at energies above $q$.
In particular, as we have already argued, the dominant contribution to 
the two-point function comes from the diagonal mode of the sites at 
energies above $q$.

Mapping the deconstructed theory back to the continuum, we see that we 
have introduced an IR cutoff on the theory, and the diagonal mode of 
the sites above $q$ maps to the zero mode of the theory with this IR 
brane in place. Thus, while one cannot safely talk about couplings of 
the zero mode of the complete theory at high energies (because of strong 
couplings in the IR) one can reasonably talk about the zero mode of the 
theory with a different IR cutoff.
It is notable that the deconstructed picture forces this 
regularization on us, and we need not appeal to holographic intuition.

Lastly, we note that these regularizations should also appear 
automatically in other, non-trivial geometries. In situations without an 
obvious regularization, the deconstructed approach, with VEVs following 
the warp factor, should give the appropriate position dependent 
regularization.

\section{Conclusions}
Extra dimensional theories with TeV-scale cutoffs invite careful 
analysis of questions regarding the ultra-violet. In the AdS slice of 
RS1, the introduction of gauge fields into the bulk offers the 
possibility of discussing short distance physics, such as coupling 
constant unification.

The application of deconstruction to this question obviates many of the 
questions raised in continuum theory analyses. Position dependent 
regulators appear automatically in the deconstructed framework, 
motivating their use in studying continuum theories.
\vskip 0.25in
{\bf \noindent Note added:} While this work was being prepared, \cite{Falkowski:2002cm} appeared which addresses similar issues. The authors would also like to acknowledge unpublished work on this subject by N.~Arkani-Hamed and A.~Cohen.

\vskip 0.15in
{\noindent \bf Acknowledgements}
\vskip 0.125in
\noindent The authors thanks Kaustubh Agashe, Antonio Delgado, 
Walter Goldberger, Witek Skiba, Veronica Sanz, Erich Popptiz and especially Yuri Shirman
 and Nima Arkani-Hamed for useful discussions.  
Y.S. and N.W. would like to thank the visitor program at 
Harvard University where this work was initiated, and the Aspen Center 
for Physics where much of this work was completed.


\end{document}